\documentclass[preprint2]{aastex}

\shorttitle{Mitigation of Pulsed Interference}
\shortauthors{Fisher et al.}

\begin{document}

\title{Mitigation of Pulsed Interference to Redshifted HI and OH Observations
  between 960 and 1215 MHz\altaffilmark{2}}

\author{J. R. Fisher} \affil{National Radio Astronomy
  Observatory\altaffilmark{1}, Green Bank, WV 24944}
\email{rfisher@nrao.edu}

\author{Q. Zhang, Y. Zheng, S. G. Wilson}
\affil{Electrical and Computer Engineering Dept., University of
  Virginia, Charlottesville, VA 22904}
\email{qingzhang@virginia.edu, yz6n@virginia.edu, sgw@ee.virginia.edu}

\and

\author{R. F. Bradley}
\affil{National Radio Astronomy Observatory\altaffilmark{1}, 2551 Ivy Rd, Charlottesville, VA 22903}
\email{rbradley@nrao.edu}

\altaffiltext{1}{The National Radio Astronomy Observatory is a facility
of the National Science Foundation operated under cooperative
agreement by Associated Universities, Inc.}

\altaffiltext{2}{A preliminary version of this work
  appeared in the NRAO Electronics Division Internal Report No. 313
  which contains additional experimental detail.}

\begin{abstract}
The neutral hydrogen 21-cm spectral line (1420.4 MHz) and the four
18-cm lines of the hydroxyl molecule (1612-1720 MHz) are observable at
redshifts which put their measured line frequencies well below their
protected frequency bands.  Part of the redshift ranges (z =
0.171-0.477 for HI and z = 0.37-0.73 for OH) fall in the 960 to 1215
MHz band that is allocated to aircraft navigation.  Most of the
signals in this band are pulsed emissions of low duty cycle so much of
the time between pulses is interference free.  This paper outlines the
structure and measured properties of signals in this band and
demonstrates a signal processing strategy that is effective at
removing the pulsed signals from spectra at sensitivities produced by
several hours of integration.
\end{abstract}


\section{Introduction}\label{intro}

Quasars emit strong continuum radiation at centimeter radio
wavelengths.  Since these objects are at cosmological distances their
radiation must past through clouds of neutral or partially ionized
hydrogen in intergalactic space and in galaxies along the line of site
to the quasar.  The hydrogen clouds selectively absorb radiation at
the 21-cm (1420.4058 MHz) ground state transition, but the observed
spectral line is Doppler shifted due to the recessional velocity of
the absorbing cloud.  The first high-redshift 21-cm absorption line
was discovered by \citet{br78} in the quasar 3C286 at 839.4 MHz (z =
0.692).  Since then a considerable number of redshifted 21-cm neutral
hydrogen (HI) absorption lines have been found
\citep[e.g.,][]{bwldt89,lane98,briggs99,lb01,kc01a,kc01b} to
frequencies as low as 500 MHz and below in both blind searches of
quasar spectra and at the redshifts of saturated Lyman-$\alpha$
absorption \citep[see][]{rauch98}.

Another spectral line emitter and absorber that can be detected at
quite large redshifts is the hydroxyl molecule, OH.  This molecule has
four lines at 1612.231, 1665.402, 1667.359, and 1720.530 MHz, of which
the two at 1665 and 1667 MHz are generally the strongest.  Emissions
from all of these lines can be very much enhanced by maser pumping
from strong far-infrared radiation and stimulated emission caused by
background continuum radiation along the line of sight (Baan, 1989).
Galaxies in early stages of nuclear evolution provide both the
far-infrared and continuum emission to produce extremely bright OH
line emissions.  Emitters of this sort are called OH megamasers
because they are roughly a million times stronger than a typical
source of OH radiation in our own Galaxy.  The most distant reported
megamaser has a redshift of z = 0.2655 ($\sim$1276 MHz) \citep{brfah92},
and a growing number of detections have been reported out to a
redshift of z = 0.2 \citep{bsl98,dg02a}.  Based on the observed
line strengths, OH megamasers should be detectable at much greater
redshifts (lower frequencies) \citep{dg02b}.

OH absorption can be seen in front of a strong continuum radio source,
such as a quasar or radio galaxy, when there is insufficient
far-infrared radiation near the OH gas cloud to pump maser emission.
Like neutral hydrogen absorption, OH absorption may be seen at any
redshift since the strength of the spectral line depends only on the
flux density of the continuum source and the optical depth of the
molecular gas.  By studying the strengths of OH emissions and
absorptions at a wide range of redshifts we can sample the physical
conditions associated with infrared emission and OH cloud temperature
and velocity structure throughout the history of the universe.  These
redshifted spectral lines are also important to the study of the
evolution of the fine structure constant over the age of the universe
\citep{darling03}.

Redshifted HI and OH lines as well as the study of pulsars and other
non-thermal physics in the universe make the radio spectrum below 1.5
GHz of considerable astrophysical importance.  This paper describes
the study of man-made radio signals in the part of this spectrum that
is allocated to aircraft navigation (960-1215 MHz) and the
demonstration of a technique for removing these signals from radio
astronomical measurements.  Transmissions from aircraft clearly
dominate this frequency range.  All of the measurements described were
made at the National Radio Astronomy Observatory in Green Bank, WV,
and the astronomical observations were taken with the 100-meter
telescope (GBT).  The signal processing algorithms applied to the data
are described in a companion paper \citep{zzwfb05} referred to as
Paper I.

\section{Spectrum Use}\label{spectrum_use}

\begin{figure}
\epsscale{.95}
\plotone{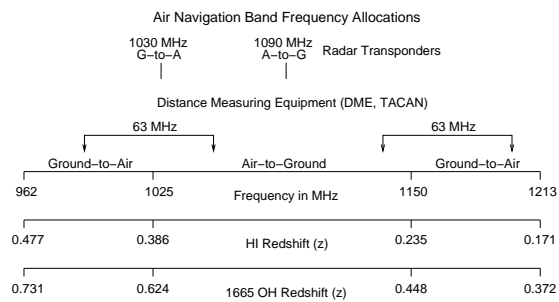}
\caption{Frequency allocations in the aircraft radio navigation band
  between 960 and 1215 MHz.}.\label{alloc}
\end{figure}

Figure~\ref{alloc} shows a graphical depiction of spectrum use between
960 and 1215 MHz along with the corresponding redshift ranges for HI
and the OH spectral lines near 1615 MHz.  This frequency range is
allocated to aircraft radio navigation, mainly used by civilian
Distance Measuring Equipment (DME) and a compatible military Tactical
Air Navigation (TACAN) system.  The air-to-ground transmissions are
confined to the 1025 to 1150 MHz range, and the 960-1024 and 1151-1215
Mhz bands are for the ground to air portion of this service.  Sharing
this band is the radar transponder system which uses 1030 MHz for
ground-to-air interrogation and 1090 MHz for air-to-ground responses.
Since many aircraft transmitting in the 1025-1150 MHz band are
line-of-sight to Green Bank, this frequency range is quite full of
strong signals during most of the day.

The DME system uses a time delay method for measuring the distance
from an aircraft to a ground station.  The aircraft begins determining
its distance by transmitting short pulse pairs at a maximum rate of 150
pulse pairs per second on the receive frequency of the selected ground
station.  After a fixed 50 microsecond delay from the received pulse
time the ground station transmits a pulse pair back to the aircraft on
a frequency either 63 MHz higher or lower than the aircraft
transmission frequency.  By measuring the delay between transmitted
and received pulses, less 50 microseconds, the aircraft's DME unit can
determine it's distance from the ground station.  This gives the
slant-range distance rather than horizontal distance, but the latter
can be computed from knowledge of the aircraft's altitude and the
elevation of the ground station.  Once the aircraft has established a
steady dialog with the ground station it slows its pulse rate to a
maximum of 30 pairs per second.  Since up to 100 aircraft can
simultaneously use the same ground station on the same frequency, the
aircraft transmitters jitter their pulse transmission intervals to
avoid locking onto ground station pulses intended for other aircraft.
See
\citet{pak97}\footnote{http://scholar.lib.vt.edu/theses/available/etd-112516142975720/unrestricted}
and \citet{bf91} for more details.

TACAN uses the same pulse timing structure as DME, but it adds azimuth
information to the ground-transmitted pulse powers.  The
civilian DME ground stations are generally co-located with VHF
Omni-Range (VOR) stations, which provide the azimuth information.  VOR
operates in the 108 to 118 MHz band.

\section{DME Signal Characteristics}\label{sig_char}

DME transmission frequencies are 1 MHz apart.  There are 126 aircraft
transmission channels running from channel 1 at 1025 MHz through
channel 126 at 1150 MHz.  There are two transmission modes for each
channel.  In mode X the ground station transmission frequency is 63
MHz below the aircraft transmission frequency for channels 1-63 and 63
MHz above the aircraft frequency for channels 64-126.  In mode Y the
frequency spacing is reversed, i.e., the ground frequencies for
channels 1-63 and 64-126 are 63 MHz above and below the aircraft
frequencies, respectively.  Hence, in mode Y the ground station
transmissions are in the air-to-ground band.  To avoid confusion
between transmissions from aircraft and ground stations on the same
frequency and to discriminate pulses from random interference, pulses
are always transmitted in pairs.  From the aircraft in mode X the
pulses are 12 microseconds apart, and in mode Y they are 36
microseconds apart.  From the ground transmitter the pulses are 12 and
30 microseconds apart in modes X and Y, respectively.  Mode X is used
much more frequently than mode Y.

The transmitted pulses are approximately gaussian in shape as a
function of time with a half-amplitude(voltage)-full-width of 3.5
microseconds.
\begin{equation}\label{pprof}
V(t) \propto e^{-0.5 (t / \sigma)^2} = e^{-2.7726 (t / W)^2}
\end{equation}
where $t$ is time, and $W$ is the full-width-half-maximum pulse width
in the same units as $t$.  Hence, from page 130 of \citet{rnb86} the
individual pulse frequency spectrum will then be gaussian in shape
\begin{equation}\label{pspec}
V(f) \propto e^{-0.5 (\omega \sigma)^2} = e^{-3.5597 (f W)^2}
\end{equation}
where $f$ is frequency in units of inverse $W$.  The voltage spectrum
full-width-half-maximum is then $0.8825 / W$, or 0.252 MHz.  The power
spectrum full-width-half-maximum is $1 / \sqrt{2}$ times that, or
0.178 MHz.

The peak pulse power from a transmitter on a large jet aircraft is 300
watts.  Manufacturers of aircraft transceivers claim a useful range of
about 550 km (300 nautical
miles)\footnote{http://www.rockwellcollins.com/ecat/br/DME-4000.html?smenu=103},
but aircraft altitude and the separations between ground stations on
the same frequency will usually limit the range to less than 300 km.
Smaller aircraft use peak powers on the order of 50 watts, which
limits their ideal range to about 250 km.  Smaller aircraft fly at
lower altitudes and have a shorter line-of sight distance to the
ground station.  An aircraft at 1700 meters (5000 feet) altitude can
expect a range of less than 100 km.  Depending on the intended range
of ground stations, the radii of their ``standard service
volumes''\footnote{http://www.flightsimaviation.com/index.php?p=\\aviationtheory\&ch=6}
range from less than 25 to 130 nautical miles (46 to 240 km) (see
Figures~\ref{all_dme} and \ref{dme_sta_121}), although
the airborne equipment is capable of longer useful ranges.

As described in Section~\ref{puls_det}, the peak power threshold for
detecting a single pulse, with a moderate number of false detections,
is about ten times the average noise power in a 0.25 MHz receiver
bandpass centered on the pulse carrier frequency.  This threshold
corresponds to a power of $6.9 \times 10^{-16}$ watts, or -151.6 dBW
at the GBT receiver input with a noise temperature of 20 Kelvins.  If
we assume a transmitter power of 300 watts, a transmitting antenna
gain of 0 dBi (decibels below isotropic gain), and a GBT sidelobe gain
of -15 dBi, the maximum free-space distance at which a single pulse
can be detected is about 2500 km.  The detectable range of a 50-watt
pulse is about 1000 km.  Neglecting atmospheric refraction, the
zero-degree horizon line-of-sight distance to an aircraft at 35,000
feet altitude is about 360 km so any aircraft DME transmitter whose
line of sight to the GBT is above the horizon will generally be
detectable in the GBT output with a signal filter matched to the pulse
bandwidth.  An aircraft with a 300-watt transmitter at 500 km distance
would be detectable below the horizon if its diffraction loss is less
than 14 dB.

As a comparison, the quoted receiver sensitivity of a representative
DME ground station
system\footnote{http://www.amsjv.com/publications/ASI11181119SLqx.pdf}
is -117 dBW for 70\% replies, and its minimum antenna gain is +8 dBi.
If we again assume a 0 dBi transmitting antenna and 300 watts
transmitted power, this corresponds to a useful line-of-sight range of
660 km.  This is very close to the claimed useful range of a 300 watt
aircraft transceiver.

\section{Measured Pulse Properties}\label{meas_pulse}

\begin{figure}
\includegraphics[angle=270,scale=0.32]{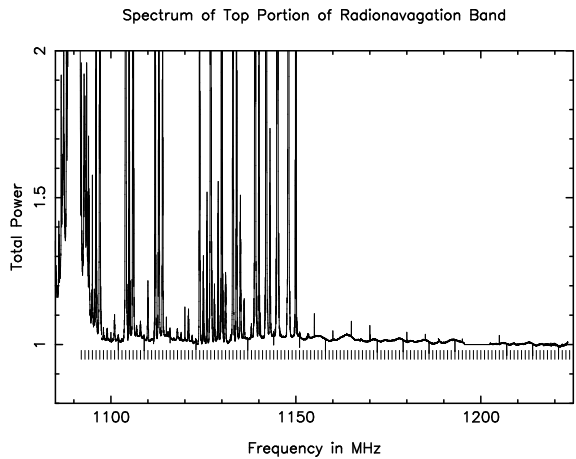}
\caption{Composite spectrum measured on the GBT of the top half of the
  aircraft navigation band (DME Channels 65 to 126), February 20, 2004
  between 12:00 and 14:00 EST.  Each 7-Mhz spectrum segment is a
  5-minute integration normalized to the receiver noise power.  The
  segment near 1200 MHz was used as a reference spectrum so it is
  shown as a flat line.  Hash marks below the spectrum show the DME
  1-MHz channel frequencies.  The vertical scale is fraction of system
  noise power.}\label{dme_spec_all}
\end{figure}

Figure~\ref{dme_spec_all} shows the composite spectrum seen by the GBT
in the frequency range of 1085-1222 MHz at the low end of the tuning
range of the 1.15-1.72 GHz, L-band receiver.  It covers DME channels
65 through 126 and the frequencies of their corresponding mode X
ground transmitters above 1150 MHz.  Signals can be seen in every DME
channel except below 1096 MHz where the radar transponder reply
signals at 1090 MHz dominate the spectrum.  A careful search for
pulses from ground stations above 1150 MHz showed no detectable DME
signals in these data.  The top trace in Figure~\ref{scan26off01tp}
shows an expanded portion of the spectrum between 1142 and 1152 MHz.

\begin{figure}
\includegraphics[angle=270,scale=0.32]{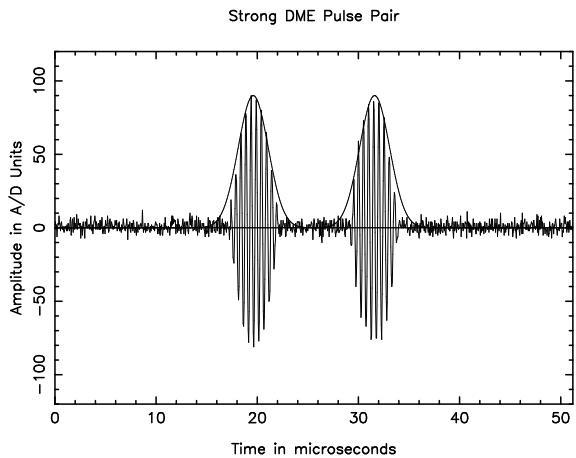}
\caption{Strong pulse pair measured at 1142 MHz.  The waveform is
slightly saturated on negative peaks.  The gaussian curves are 3.5
microseconds half-amplitude width and 12 microseconds
apart.}\label{pulse_pair}
\end{figure}

Figure~\ref{pulse_pair} shows the waveform of a pair of strong pulses.
These pulses nearly saturated the RF amplifiers so the negative peaks
are slightly compressed.  Plotted over the pulses are two gaussians
with 3.5 microsecond full-width-half amplitude spaced 12 microseconds
apart.  This particular transmitter appears to truncate its pulses
somewhat faster than a gaussian curve, hence the spectral bandwidth
will be somewhat wider than the values stated in
Section~\ref{sig_char}.  Also, the pulses are a couple of tenths of a
microsecond closer together than 12 microseconds.  This may be typical
of the tolerances to be expected.

\begin{figure}
\includegraphics[angle=270,scale=0.32]{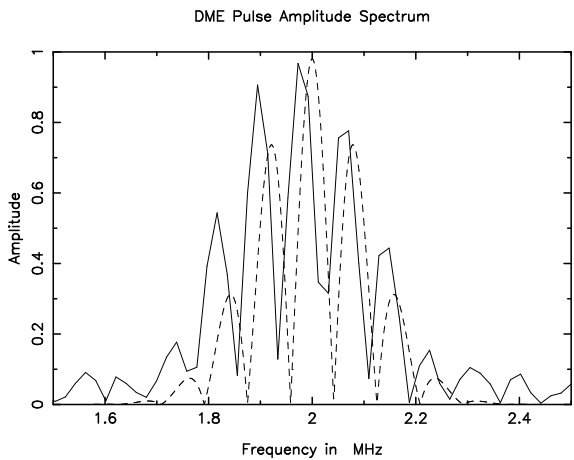}
\caption{Amplitude spectrum of the pulse pair in
Figure~\ref{pulse_pair}.  The dashed curve is the spectrum
expected from a pair of gaussian pulses of 3.5 microseconds amplitude
width and 12 microseconds apart with a carrier frequency of 1142.000
MHz.}\label{pulse_spec}
\end{figure}

Figure~\ref{pulse_spec} shows the Fourier transformed amplitude
(voltage) spectrum of the pulses shown in Figure~\ref{pulse_pair}.  As
expected from the measured pulse shape, the amplitude spectrum is
wider than would be expected from a gaussian pulse shape of 3.5
microseconds half-amplitude width.  Also, the carrier frequency is
about 21 kHz lower than the nominal 1142.0 MHz of this transmitter
channel.  (A manufacturer's DME transceiver
brochure\footnote{http://www.rockwellcollins.com/ecat/br/DME-42\_442.html?smenu=103}
states a frequency stability specification of $\pm 100$ kHz.)  The 83
kHz modulation period in this spectrum is due to coherent beating of
the two pulses.  In Paper I the use of matched filters
for optimum pulse detection is explained, and the frequency offset
observed here shows that the filter must match single pulses rather
than pulse pairs to avoid missing pulses that fall in the nulls of a
two-pulse matched filter.

\begin{figure}
\includegraphics[angle=270,scale=0.32]{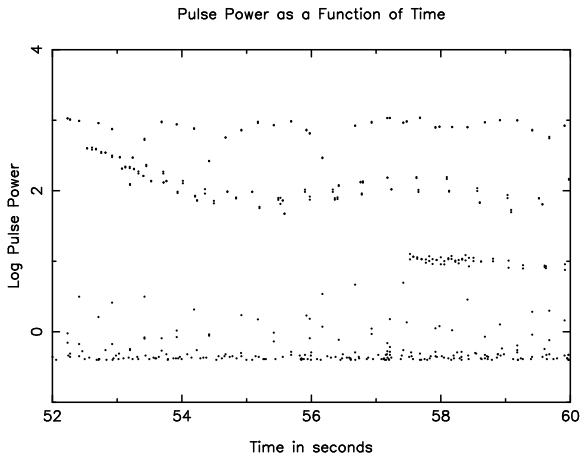}
\caption{Measured powers of individual pulses vs time at 1145 MHz (DME
  Channel 121) recorded 5:53 EST on January 27, 2004 with the GBT
  L-band receiver.  Individual pulses in each pulse pair are
  shown.}\label{pulse_train}
\end{figure}

Figure~\ref{pulse_train} shows a fairly typical train of pulses from
several aircraft.  There appear to be at least five
transmitting aircraft in this plot as surmised from the tracks of
pulse power as a function of time.  Three of the transmissions
began in the time interval shown at about 52.5, 53.0, and 57.5
seconds.  Each began with a repetition rate of about 20 pulses per
second and then slowed to about 4 or 5 pps.  This is a considerably
slower pulse rate than the 120-150 and 24-30 pps acquisition and
post-acquisition rates described in the literature \citep{bf91}.  One
manufacturer's product
brochure\footnote{http://www.rockwellcollins.com/ecat/br/DME-42\_442.html?smenu=103}
states that it is capable of scanning three DME channels at 12.5
milliseconds per channel so that the DME samples each channel 27 times
each second. Even when covering three channels at once, it can lock-on
in less than one second.  This fits the observed behavior of the three
pulse tracks in Figure~\ref{pulse_train}.  After acquisition the
transceiver appears to scan the channels at a slower rate, which is
advantageous from radio astronomy's point of view since the average
power received from each aircraft transmitter is lower at a given
frequency.

The detection threshold used in generating Figure~\ref{pulse_train} is
about ten times the average receiver noise power in the 0.25 MHz
bandwidth of the pulse.  In Section~\ref{sig_char} we calculated this
threshold to be -151.6 dBW at the GBT receiver input with a noise
temperature of 20 Kelvins.  The stated receiver sensitivity of a
commercial DME ground station is -117
dBW~\footnote{http://www.amsjv.com/publications/ASI11181119SLqx.pdf}.
If we assume that the system noise temperature of the DME station is
400 K, then the GBT detection threshold used is about 21.6 dB = $151.6
- 117 - 10 * Log_{10}(400/20)$ lower in terms of system noise power
than is used in the DME station receiver.  The DME ground station
detection threshold would correspond to a level of about 1.8 on the
vertical scale of Figure~\ref{pulse_train}.  We surmise that this
threshold difference is roughly the signal processing margin used in
the DME station receiver to avoid false detections or detections of
distant aircraft interrogating other stations on the same frequency.

\section{Ground Station Locations}\label{sta_loc}

\begin{figure}
\includegraphics[angle=270,scale=0.32]{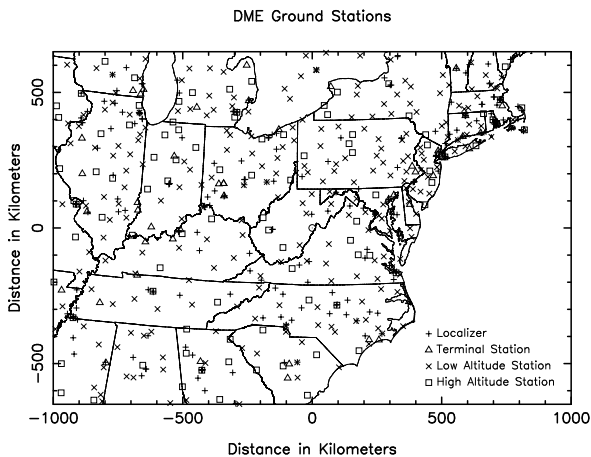}
\caption{The locations of all DME stations around Green Bank found in
the Navaid Digital Data File of the Digital Aeronautical Information
CD data base from the National Aeronautical Charting Office of the
Federal Aviation Administration.  The high altitude stations are
marked with squares, the low altitude stations are marked with X's,
the terminal stations are marked with triangles, and localizer
stations are marked with +'s.  The location of Green Bank is marked
with a circled dot at the coordinates (0,0).}\label{all_dme}
\end{figure}

Figure~\ref{all_dme} shows the location of DME stations in the
eastern U. S. that are found in the aeronautical information data
base.\footnote{http://www.naco.faa.gov/index.asp?xml=naco/catalog/\\charts/digital/daicd}  Stations with different service radii and altitude ranges are
shown with different symbols.

\begin{figure}
\includegraphics[angle=270,scale=0.32]{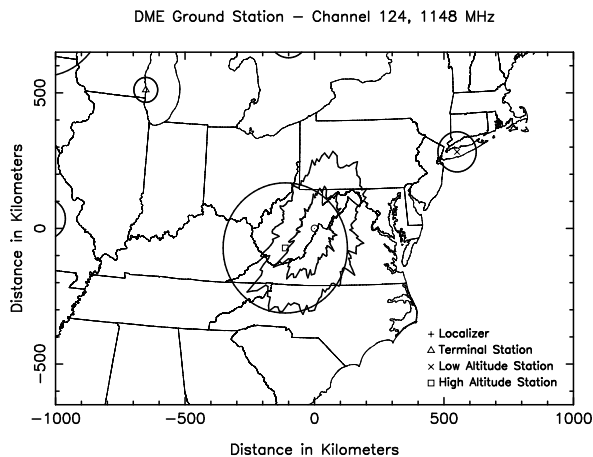}
\caption{The locations of DME channel 124 (1148 MHz) stations.  The
  range of each station is shown by the circle centered on it: 240 km
  for high altitude stations, 74 km for low altitude, and 46 km for
  terminal stations.  No localizer stations are shown for this
  channel, but they would be surrounded by 20 km radius circles.  The
  irregular contours around Green Bank, WV (indicated by the circled
  dot) show the line-of-sight horizons of the GBT prime focus point
  for aircraft at altitudes of 4.6 (inner) and 10.7 (outer) km (15,000
  and 35,000 feet).}\label{dme_sta_121}
\end{figure}

An example of the locations of stations operating in a specific DME
channel is shown in Figure~\ref{dme_sta_121}.  The large circles
illustrate the greater range of the high-altitude stations.  Around
the location of Green Bank two irregular contours show the
line-of-sight distance to an aircraft just above the horizon as seen
from the prime focal point of the GBT for aircraft altitudes of 15,000
and 35,000 feet above sea level.  The horizon contours do not take
atmospheric refraction into account, but refraction extends the
effective line-of-sight range by only 5 to 10\%.

\section{Pulse Detection and Statistics}\label{puls_det}

To generate a test data set under conditions close to typical GBT
observing a series of on-off scans were taken on the radio source
0952+176 which has a continuum flux density of about 1.4 Janskys and a
narrow HI absorption line at 1147.5 MHz with a depth of about 15 mJy
\citep{kc01b}.  These observations used 8-bit baseband sampling of a
10 MHz bandwidth centered on 1147 MHz.  This passband included DME
channels 119 (1141 MHz) through 126 (1150 MHz) plus channel 118 at the
lower edge of the band and a bit of spectrum above the air-to-ground
frequency range.  Two observing sessions were recorded: January 27,
2004 from 05:24 to 07:37 EST (13:29 to 15:42 LST) and February 23,
2004 from 17:32 to 20:01 EST (03:35 to 05:54 LST).  Unless otherwise
noted, all of the pulse property measurements used the January data.
The February session was used to verify the efficacy of pulse blanking
described in Section~\ref{blanking}

These data were processed in one-minute chunks, which was the file
size into which the data acquisition computer divided the 8-bit
samples.  The A/D input level was such that the noise voltage rms
value spanned about four A/D levels.  The sample interval was 50
nanoseconds (20 mega-samples per second).

The first major signal processing step was to scan every one-minute
data file for pulses and create a list of pulse times and powers
for each DME channel in the spectrum.  The pulse detection scheme is
described in detail in Paper I.  The basic process was to
Fourier transform one-megasample (50 millisecond) data lengths into
the frequency domain, apply a matched filter to each DME frequency,
transform back to the time domain, square to get power, and search
this power sample series for peaks above a chosen threshold.  The
threshold was set to 15 times the median power value in the first
50-millisecond sample of each DME channel.  This was roughly 10 times
the average power level in a channel filter passband between pulses
and produced a modest number of false detections on random noise.
Pulse peaks were then found starting with the strongest one.  Data in
a 15-microsecond window centered on the found peak were set to zero so
that this pulse and surrounding data would not be detected in the
remainder of the peak search.  This window was small enough to allow
the second pulse in a 12-microsecond pair to be detected but wide
enough to suppress most of the wings of strong pulses.  The final
pulse list was sorted by pulse time for further analysis and for use
in blanking pulses in the original data set.

During the analysis of pulse statistics the strongest pulses were
found to affect the entire 10-MHz spectrum causing false detections in
channels other than the one containing the transmitted pulse.  To
remove this spurious effect from the analysis the sorted pulse lists
were searched for strong pulses, and all pulses in other channels
within five microseconds of each strong pulse were deleted from the
list.

The filter applied to each DME channel frequency was chosen by
visually matching the envelope of the spectrum shown in
Figure~\ref{pulse_spec} to a gaussian function.  This function had a
full-width-half-maximum width of 0.25 MHz in the voltage domain.  This
corresponds to a power noise equivalent width of 0.188 MHz.  If a DME
pulse transmitter were as much a 0.1 MHz off in frequency, as is
permitted in at least one manufacturer's specs, the filter response
would be down by about 4 dB, but most of the pulses appeared to be
within about 30 kHz of the nominal channel frequencies.

\begin{figure}
\includegraphics[angle=270,scale=0.32]{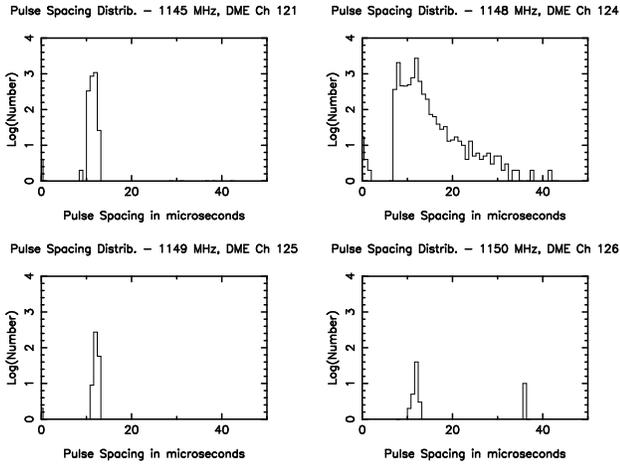}
\caption{Distribution of adjacent pulse spacings for four DME channels
measured for one minute near 06:22 EST of January 27.  The continuous
distribution of pulse spacings shown in the top right panel is
probably due to echoes from surrounding terrain as explained in the
text associated with
Figure~\ref{pulse_echoes_ch6}}\label{pulse_spacing}
\end{figure}

We expect pulses to arrive in pairs, either 12 or 36 microseconds
apart for mode X or Y, respectively.  This should be evident in a
histogram of adjacent pulse separations.  Figure~\ref{pulse_spacing}
shows four such histograms for one minute of measurements of four
different DME channels.  All four channels show a clear peak at 12
microseconds, and channel 126 shows a small number of mode Y pulses at
36 microseconds.  The nearest ground station assigned to channel 124
is in Beckley, WV so there is a lot of activity on this channel.  The
large number of pulses on channel 124 at spacings other than 12
microseconds may be due to echoes of strong pulses from terrain around
Green Bank as described in Section~\ref{echoes}.

\begin{figure}
\includegraphics[angle=270,scale=0.32]{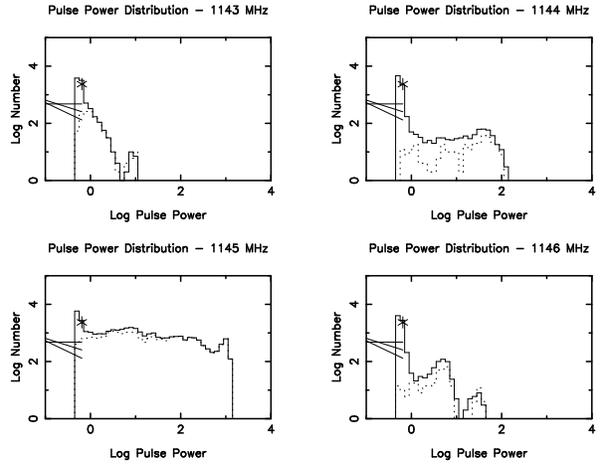}
\caption{Number distribution of detected pulses as a function of
  relative pulse power for DME channels 119-122 in a five minute scan
  recorded at 06:22 EST of January 27.  The solid lines show all
  detected pulses, and the dashed lines show only pulses that are
  paired with another between 11 and 13 microseconds away.  On the
  left side of each plot are plotted the pulse number distribution
  limits to remain undetected in a five-minute integration as
  described in the text.  The star marks the limit if all unblanked
  pulses are at the detection threshold ($Log_{10}(power) \approx -0.2$).
  The horizontal straight line is for a number density index of
  $\alpha = 0$, and the sloped lines are for $\alpha = -0.5$ and
  $-0.75$.  Pulses saturate the receiver at $Log_{10}(power) \approx
  3.0$.}\label{scan15onloH}
\end{figure}

Figure~\ref{scan15onloH} shows examples of pulse power distributions,
ranging from the detection threshold to the pulse saturation limit of
the GBT receiver for four DME channels in one five-minute scan.  The
solid and dashed lines in these plots show the difference between,
respectively, all detected pulses and only those pulses that have
detected neighbors near 12 microseconds away.  The difference gives an
upper limit on the number of false detections due to random noise or
sources of non-DME pulses.  An interesting feature of these plots is
that the number of known pulse pairs per logarithmic power interval
does not appear to be rising steeply with decreasing power near the
detection limit.  This is fairly typical of most of the data in these
two hours of GBT observations.

Two important questions to ask are how much power remains in pulses
below the detection threshold, and is this sufficient to be noticeable
in the spectrum with detected pulses removed?  To answer these
questions an assumption is needed for the number distribution of the
undetected pulses as a function of power.  Assuming a power law for
the pulse power distribution
\begin{equation}\label{log_pwr_dist}
log({{n(I)} \over {n_o}}) = \alpha~log\biggl({{I} \over {I_o}}\biggr)
\end{equation}
or
\begin{equation}\label{pwr_dist}
{{n(I)} \over {n_o}} = \biggl({{I} \over {I_o}}\biggr)^\alpha
\end{equation}
then the total power in the undetected pulses will be
\begin{equation}\label{tot_pulse_pwr}
p = \sum_i n(I_i)~I_i = \sum_in_o~\biggl({{I_i} \over {I_o}}\biggr)^\alpha~I_i
\end{equation}
If we do the summation using logarithmic intervals in $I_i$ of 0.1 for
various values of $\alpha$ we get the values of $p/I_o$ shown in the
second column of Table~\ref{undet_pulses}.  For example, for $\alpha =
0$, if $I_o$ is the pulse power detection threshold and $n_o = 1$, the
cumulative power from all pulses below the threshold will be 4.9 times
the power in one pulse at the threshold level.  For values of $\alpha
\leq -1$ the summation in Equation~\ref{tot_pulse_pwr} will diverge at
low powers, and such a distribution will give infinite power
without a low power cutoff.  Since each DME channel will be
occupied by a finite number of aircraft transmissions such a
divergence is not possible in practice.

To determine limits on the number of pulses under the detection
threshold that can be tolerated in an astronomical spectrum, assume a
five-minute integration with a spectral resolution of 20 kHz (5.2 km/s
at 1.15 GHz).  The rms noise power in each spectral channel
will be $4.1\times10^{-4} = 1 / (300^s\times 2\times10^4$ Hz) times
the total noise power in the channel.  Most pulses stronger than 15
times the average noise power in the bandwidth of the pulse will be
detected.  The width of one pulse is 3.5 microseconds so the power in
one pulse averaged over the 300 second integration is
$15\times3.5\times10^{-6}/300 = 1.75\times10^{-7}$ times the noise
power in the pulse bandwidth.  Hence, there could be
$4.1\times10^{-4}/1.75\times10^{-7} = 2343$ pulses at the detection
threshold (7.8 per second) without significantly affecting the
spectrum.  If the undetected pulse powers were distributed with a
power law of $\alpha = -0.5$, then there could be $2343/9.2=255$
pulses (0.85 per second) in the interval just below the detection
thresholds and more at weaker powers according to the power law.
The tolerable number of pulses per second in the interval just below
the detection threshold for various values of $\alpha$ is given in
the third column of Table~\ref{undet_pulses}, and the total pulse
number limits are plotted in Figure~\ref{scan15onloH}.

\begin{table}
\caption{Undetected pulse residual power and limits on the number of
  pulses per second in the interval just below the detection threshold
  in a 300-second integration for different pulse power
  distributions} \label{undet_pulses}
\begin{center}
\begin{tabular}{ccccc}
$\alpha$ & $p/I_o$ & $N(I_o)/second$ \\
\hline
+inf  & 1.0 & 7.8 \\
0     & 4.9 & 1.6 \\
-0.25 & 6.3 & 1.2 \\
-0.50 & 9.2 & 0.85 \\
-0.75 & 17.9 & 0.43 \\
\end{tabular}
\end{center}
\end{table}

The limits shown on the left side of the plots in
Figure~\ref{scan15onloH} show that the unblanked pulses are close to
being detectable in the astronomical spectra for the more active DME
channels measured.  The assumptions that went into the calculation of
these limits may be a bit pessimistic.  In an on-off observing
procedure the average pulse power will tend to cancel, and the
one-sigma power limit is probably a bit too stringent.
Nevertheless, the unblanked pulses are of some concern, and
improvements in pulse detection sensitivity may be worth pursuing.
For a given pulse power distribution, the number of tolerable
pulses below the detection threshold is inversely proportional to the
threshold power.

\section{Signal Power vs DME Station Distance}

\begin{figure}
\includegraphics[angle=270,scale=0.32]{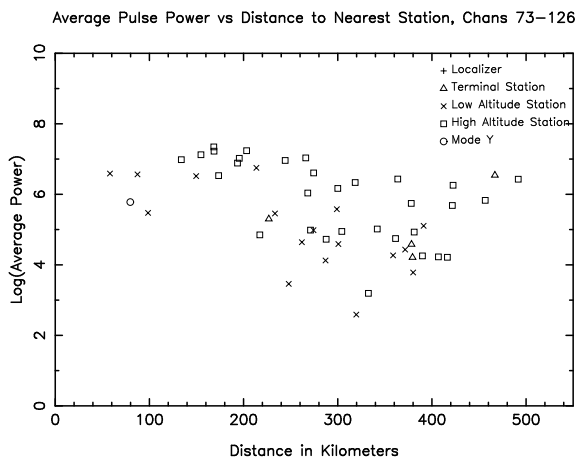}
\caption{Relative average signal power in five minutes as a function of
  the distance to the nearest DME station from Green Bank for all DME
  channels above 1100 MHz shown in Figure~\ref{dme_spec_all}.  The
  high, low, terminal and localizer station symbols are the same as in
  Figures~\ref{all_dme} and
  \ref{dme_sta_121}.}\label{pwr_vs_short_distance}
\end{figure}

One might expect to measure lower average signal powers on DME
channels where the nearest ground station is farther away.  To test
this assumption the relative average powers of all pulses detected
were plotted as a function of nearest station distance.  The results
are shown in Figure~\ref{pwr_vs_short_distance} for all channels shown
in Figure~\ref{dme_spec_all} above 1100 MHz.

A notable feature of Figure~\ref{pwr_vs_short_distance} is that
relatively strong signals can be seen for stations out to at least 500
kilometers distance, particularly for high-altitude stations.  There
may be saturation effects in the power of signals from nearby
stations so the upper envelope of the distribution may be somewhat
suppressed at short distances.  As expected, the signals associated
with low-altitude stations are generally weaker than those associated
with high-altitude stations.  Not all signals come from aircraft that
are interrogating the nearest station so conclusions can be drawn only
from trends in this figure and not from individual data points.  There
was a slight tendency for the total pulse power to increase with time
over the observing session as one might expect in the early morning
hours as aircraft begin their flight schedules.

\section{Echoes}\label{echoes}

\begin{figure}
\includegraphics[angle=270,scale=0.32]{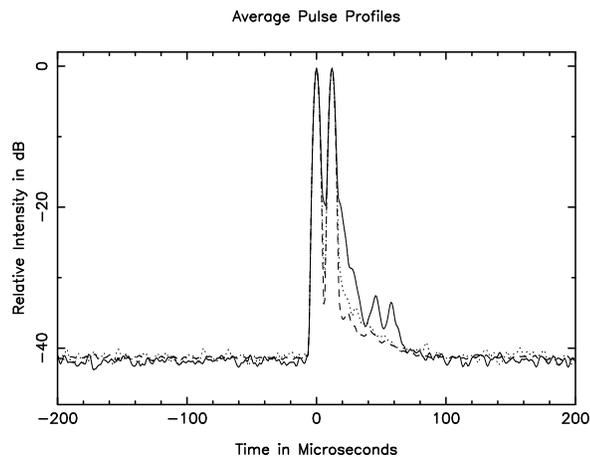}
\caption{Average pulse profiles at 1145 MHz (DME channel 121, cf. the
  top right panel in Figure~\ref{pulse_spacing}) in one minute scans
  recorded at 05:24 (solid), 07:06 (dashed), and 07:27 (dotted) EST.
  Only isolated pulses in a 2-dB range of amplitudes just below GBT
  receiver saturation were included in the
  averages.}\label{pulse_echoes_ch6}
\end{figure}

The broad distribution of pulse spacings shown in the top right panel
of Figure~\ref{pulse_spacing} could be due to echoes of strong pulses
from surrounding terrain as were measured in pulses from a
ground-based air surveillance
radar\footnote{http://www.gb.nrao.edu/$\sim$rfisher/Radar/analysis.html}.
To test this hypothesis average pulse power profiles were computed in
a number of one-minute intervals for the strongest pulses that did not
saturate the receiver system.  The results for the DME channel in the
top right panel of Figure~\ref{pulse_spacing} are shown in
Figure~\ref{pulse_echoes_ch6}.  All of the profiles show strong
evidence of echoes out to a delay of about 70 microseconds.  This
agrees well with the measured distribution of radar pulse echo delays.
Echo pulse powers relative to the direct pulse powers range from
about -20 dB at small delays to less than -40 dB, which is the
detection limit of these measurements.  The solid trace in
Figure~\ref{pulse_echoes_ch6} shows what could be an isolated single
reflection at a delay of about 45 microseconds.

\section{Pulse Blanking in the Integrated Spectra}\label{blanking}

To remove DME signals from the data the list of detected pulse times
in each one minute record were used to set blanking windows in the raw
sampled data.  Each blanking window was 12 microseconds wide, and it
was applied to each detected pulse, including a small fraction of
false detections due to random noise.  Unlike the strategy described
in Paper I, a detected pulse in any of eight DME channels (1143-1150
MHz) resulted in blanking data for the full 10 MHz bandwidth.  In the
January 27 data this resulted in a blanked data fraction ranging from
0.5 to 1.1\% of the data samples with the fraction increasing with
time over the 05:24 to 07:37 EST observing period.  A smaller fraction
of data could be blanked if one-megahertz sections were processed
separately, blanking only on pulses from one DME channel at a
time.  However, this adds the complication of stitching the spectra
back together seamlessly, and strong pulses from adjacent channels
that splatter across the spectrum may not be blanked as effectively.
Even in the busiest part of the day the data lost probably will not be
greater than about 2\% with full 10-MHz spectrum blanking.
Simultaneous processing of wider bandwidths would lose proportionately
greater fractions of data.

One consequence of blanking small windows of sampled data is that
power is lost, and extra sidelobes will be generated on narrowband
signals.  The fraction of power lost and the total power in the
sidelobes will be equal to the blanking fraction.  One or two percent
loss usually will not affect the astronomical data significantly, but
this effect needs to be kept in mind.  In principle, one can compute
and correct for the sidelobe pattern and power loss from the known
blanking pattern, if the extra computational load is warranted.

\begin{figure}
\includegraphics[angle=270,scale=0.32]{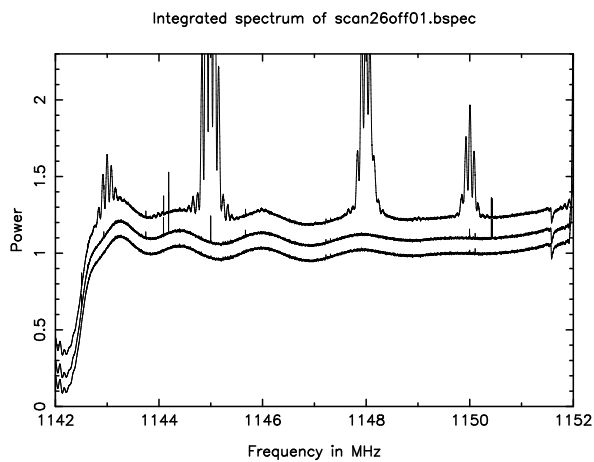}
\caption{Total power spectra computed for one minute of data recorded
  at 07:23 EST on January 27 on the GBT.  The spectra contain
  8192 channels (1.2 kHz channel spacing).  The top trace
  is without blanking.  The middle trace is with blanking, and
  the bottom trace is with blanking and narrowband signal
  removal.}\label{scan26off01tp}
\end{figure}

Figure~\ref{scan26off01tp} shows the total power spectrum of one
minute of data.  Strong DME signals can be seen in the top spectrum,
without blanking, in four of the channels, and weak signals can be
seen in at least three other channels, not including 1142 MHz.
The middle trace shows the spectrum from the same data with pulse
blanking.  To the noise level of this spectrum, all of the blanked DME
signals have been suppressed.  There are still a few RFI signals in
the spectrum, but most of these are quite narrow and can be excised by
deleting one to three spectral channels around each narrow spike as
shown in the bottom spectrum.  None of the narrow spikes are likely to
be associated with the DME signals.

To test how well pulse blanking works on long astronomical
integrations the data from the January 27 and February 23 two-hour
observing sessions of the quasar 0952+176 were run through the
blanking algorithms.  The observing sequence was alternately 5 minutes
on source and 5 minutes off.  The total on- plus off-source
integration times in the sessions were 108 and 140 minutes,
respectively.  The data were summed with weighting of each spectrum by
the inverse of the square of the estimated system temperature for each
ten-minute on-off scan pair.  The weighted system temperature
estimates in each session were 21.8 and 23.4 Kelvins, respectively.
The February session was observed at lower elevation angles.

\begin{figure}
\includegraphics[angle=270,scale=0.32]{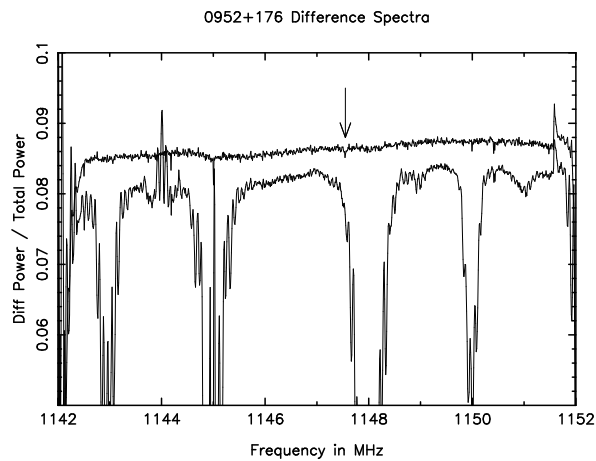}
\caption{Averaged (ON - OFF) / OFF spectra for all of the January 27
  data.  The vertical scale is in fraction of the system noise power.
  The bottom spectrum is without blanking and the top spectrum is with
  blanking and narrowband spike removal.  The arrow marks the
  frequency of the absorption line in 0952+176.}\label{integ_spec_127}
\end{figure}

Figure~\ref{integ_spec_127} shows the integrated spectra from the
first observing sessions with and without pulse blanking and
narrowband spike removal.  A similar plot for the second session
looks much the same.  The weak absorption feature at 1147.5 MHz can be seen in
the blanked spectra.  There is a hint of residual DME signal in the
strongest channel frequencies as we expect from the analysis in
Section~\ref{puls_det}, but this does not affect the detection of
relatively narrow spectral features significantly.

\begin{figure}
\includegraphics[angle=270,scale=0.32]{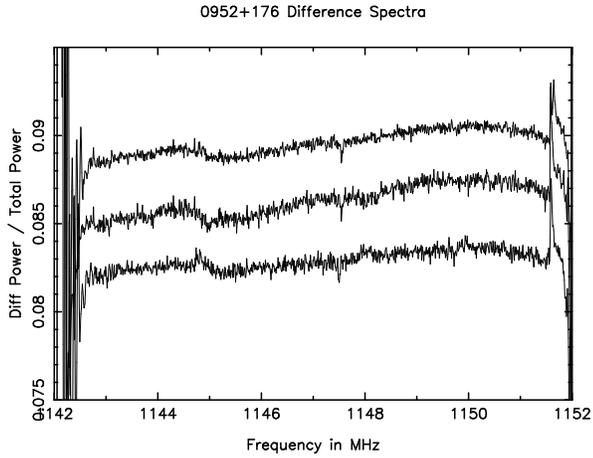}
\caption{Averaged (ON - OFF) / OFF spectra for all of the data with
  blanking and narrowband spike removal.  The middle spectrum
  is from January 27; the bottom spectrum is from February 23;
  and the top  spectrum is the average of all data with the
  February 23 spectrum shifted by the expected differential Doppler
  shift.  The spectra are offset vertically by a small amount for
  clarity.}\label{integ_spec_all}
\end{figure}

\begin{figure}
\includegraphics[angle=270,scale=0.32]{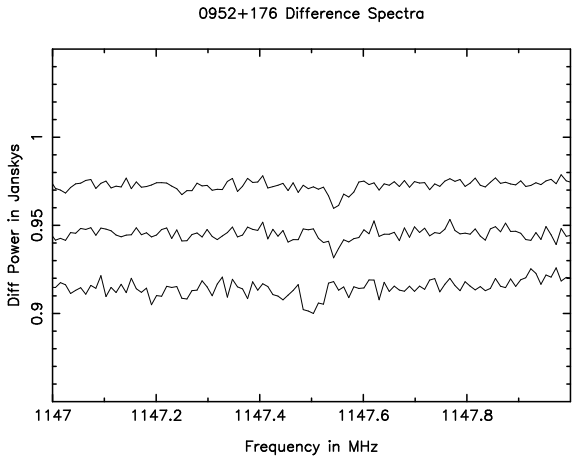}
\caption{Same as Figure~\ref{integ_spec_all} but expanded around the
  1147.5 MHz HI absorption feature and with the vertical scale in
  Janskys.  The lower two spectra are offset vertically by a small
  amount for clarity.}\label{integ_spec_all_exp}
\end{figure}

Figures~\ref{integ_spec_all} and \ref{integ_spec_all_exp} show the
integrated spectra from both observing sessions.  The expected 52 kHz
differential Doppler shift in the HI line frequency between the two
observing sessions can be seen in these plots.  The February 23
spectrum is shifted to the January 27 Doppler offset before taking the
weighted average of the two spectra.  The vertical scale in
Figure~\ref{integ_spec_all_exp} is shown approximately in Janskys.
The continuum level around 0.97 Jy is less than the published 1.4 Jy,
but the measured peak line depth to continuum ratio of $0.0127\pm0.02$
is quite close to the published value of 0.013 by \citet{kc01b}.  The
same paper shows a line half-power width of about 35 kHz, which is
consistent with the line widths shown in
Figure~\ref{integ_spec_all_exp}.

The measured single-channel rms noise values of the three spectra
shown in Figure~\ref{integ_spec_all_exp}, without any baseline
removal, are 3.7, 3.2, and 2.8 milli-Janskys, respectively, bottom to
top.  From the resolution bandwidth and integration times and the
known correlation loss due to the way the spectra are generated in the
FFT algorithm the expected rms values are 3.6, 3.2, and 2.4
milli-Janskys, respectively.  Hence, the pulse blanking scheme
described in this paper is fully effective with minimal loss of data
and little, if any, loss in sensitivity.

\section{Computational Requirements and Future Work}\label{comp}

All of the signal processing described in this report was done on a
general purpose processor.  Implementation of faster algorithms
remains to be done, but it is clear from a rough analysis of the
computational load that finding DME pulses in the data and computing
blanked spectra for more than a few megahertz of bandwidth will
require a fairly large computer cluster or one of the parallel
processing digital technologies.  We are investigating the feasibility
of programming the critical algorithms into field programmable gate
arrays (FPGAs).  Much of the science to be accomplished with the
technique described in this paper and Paper I will require real-time
processing of more than 100 MHz of bandwidth.

More sensitive pulse detection will help remove weaker pulses that are
probably leaking into the pulse-blanked spectra shown in this paper.
The filter used for pulse detection is reasonably close to optimum so
the only route for weaker pulse detection is an antenna and RF
amplifier system that has a higher signal-to-noise ratio on DME
signals.  The GBT has an approximate average sidelobe gain of -15 dBi
and a system temperature of 20 Kelvins.  If a separate antenna with a
gain of +6 dBi and a pre-amplifier that produces a system
temperature of 400 Kelvin were used to receive the DME signals, then
one could realize an increase in SNR of about 8 dB = 6 dBi - (-15 dBi)
+ 10 log(20K/400K).  Implementation of a receiving system for this
purpose is now underway.

\acknowledgments

Facilities: \facility{NRAO(GBT)}.

\end{document}